# REGULAR AND CHAOTIC COHERENT STATE DYNAMICS OF SEVERAL QUANTUM OPTICAL MODELS

Alexander V. Gorokhov, Elena V. Rogacheva, and Alexander V. Shiryaev

*Department of General and Theoretical Physics,*

*Samara State University,*

*ul. akadem. Pavlova 1, Samara, 443011, Russia,*

*Phone: (8462)345434, FAX: (8462)345417,*

*e-mail: gorokhov@info.ssu.samara.ru*

## Abstract

The coherent state representations of the group $G = W_1 \otimes G_0$ (where $G_0 = SU(2), SU(1,1)$) are used in computer simulation of the dynamics of single two-level atom ($G_0 = SU(2)$) interacting with a quantized photon cavity mode - the Jaynes - Cummings model (JCM) without the rotating wave approximation and, in general, nonlinear in photon variables). The second case (hyperbolic Jaynes - Cummings model (HJCM), $G_0 = SU(1,1)$) corresponds to the quantum dynamics of quadratic nonlinear coupled oscillators (the parametric resonance on double field frequency and a three - wave parametric processes of nonlinear optics). Quasiclassical dynamical equations for parameters of approximately factorizable coherent states for these models are derived and regimes of motion for "atom" and field variables are analyzed.

# I. INTRODUCTION

Classical chaos as an unpredictible motion in systems with causal dynamics is well understood phenomenon [1–3]. Many approaches have been suggested during the past decade to define chaos in quantum systems. Nevertheless, this problem is far from clear [6,4]. We will use in this work the terminology of Refs. [5,6], where quantum chaos has been defined as the study of semiclassical behaviour of parameters of quantum system whose classical analog exhibits chaos.

Very interesting models in the study of quantum chaos properties are models of quantum optics, among which are most important ones describing interaction between atoms and light [7]. The Jaynes - Cummings model (JCM) [8] of a single two - level atom interacting with a single - mode cavity photon field attracted widespread attention of many investigators [9–11]. It has been observed theoretically [12,13] that the model possesses a considerable and unexpected pure quantum features (periodic collapses and revivals of the mean atomic exitation energy and the mean photon numbers).

Belobrov, Zaslavsky and Tartakovsky [14] first studied the dynamics of the many - atom JCM at the semiclassical approximation. Later several variants of this model have been investigated by different researchers [15,7,16,17]. It was found that the chaos came into view in this system when the rotating wave approximation (RWA) is not used, and this implies the very large values of transition dipole moments and atom densities. Gorokhov and Ruchkov [18] was observed that semiclassical chaos may be efficiently suppressed when possible parametric instabilities of nonlinear media packed in the cavity are taken into account.

The possibility of chaotic behaviour for system which consists of the single two-level atom coupling with the one - mode of quantized electromagnetical field in ideal resonator, is also discussed at the last time [19]. Contrary to the many - atom systems in this situation the ratio of atom - field coupling constant to transition frequency is usually very small and chaotic regime cannot appears.

Up-to-date technological achievments [23] lead to practical interest for systems with coupling constant of order of frequency $\omega$, when RWA can not be applied.

The experiments with the one-atom maser [20,21] and recent realization of an optical microlaser [22,23] gives hope to achieve more favourable values of this ratio and to observe chaos for this fundamental quantum - electrodynamical system.

Besides, it is interesting to consider different quantum optical models with the aim of checking their possible quantum chaotic properties. The two - level atom with $m-$ quantum transitions should be studied first. Another case is a quantum model of the three - wave parametric interaction [24]. As we show here, it leads to hyperbolic JCM - a system with model hamiltonian in terms of $SU(1,1)$ group generators instead of the Pauli matrices describing two - level atom.

The paper is organized as follows. Section 2 presents a brief description of model hamiltonians and basic assumptions. Section 3 is devoted to the introduction of relevant group theoretical coherent states (CS) and their applications to the derivation of motion equations using CS path integral representation for a transition amplitude. The most important feature of coherent states is thatthey are pure quantum states, which are the most close for classical ones [25,26] . The coherent-state point's motion obeys classical-like dynamical equations, and it is reasonable to use CS basis for invoking chaotic mystery into quantum mechanics. Section 4 deals with the results of computer simulation of systems discussed. In Conclusions we have considered possible generalizations of model hamiltonians and developing approaches for purpose of taking into account quantum corrections and their influence on chaos.

## II. MODEL HAMILTONIANS

Let us consider the two - level atom model and some of its generalizations. It is well known, that the quantum hamiltonian for two-level atom, coupling one mode of quantized electromagnetical field in an ideal cavity, can be represented as the following [11]:

$$\mathbf{H} = \mathbf{H}_F + \mathbf{H}_A + (\,\mathbf{H}^{(1)}_{int} + \mathbf{H}^{(2)}_{int}\,)\,, \tag{2.1}$$

$$\mathbf{H}_F = \nu\,\left(\,\mathbf{b}^+\mathbf{b} + \frac{1}{2}\,\right)\,, \tag{2.2}$$

$$\mathbf{H}_A = \omega\,\mathbf{J}_0\,, \tag{2.3}$$

$$\mathbf{H}^{(1)}_{int} = g\,\mathbf{b}\mathbf{J}_+ + g\,\mathbf{b}^+\mathbf{J}_+ + h.c.\,, \tag{2.4}$$

$$\mathbf{H}^{(2)}_{int} = k\,(\,\mathbf{b}^+ + \mathbf{b}\,)^2\,, \tag{2.5}$$

here $\nu$ is (angular) frequency of the field, $\omega$ is atomic frequency and difference between its energy levels, as we choose $\hbar = 1$, $g$ and $k$ are coupling constants, $\mathbf{b}^+$ and $\mathbf{b}$ are the boson creation and annihilation operators, $\mathbf{J}_0$, $\mathbf{J}_\pm = \mathbf{J}_1 \pm \mathbf{J}_2$ are the generators of $SU(2)-$ group, which equal one half of corresponding Pauli matrices (only two levels are taken into account and operators $\mathbf{J}_\pm$ increase (decrease) the energy of the atom by $\omega$ ).

In agreement with the model, $\nu \approx \omega$ and the resonant terms in the Hamiltonian $\mathbf{H}^{(1)}_{int}$ give principal contribution, when amplitude of transitions is calculated in the case of low values of constant $g$. Usually, the term $\mathbf{H}^{(2)}_{int}$ is eliminated, because $k \ll g$. Also, $g\,\mathbf{b}^+\mathbf{J}_+$ and hermitian conjugate term discard in rotating wave approximation (RWA). Finally, Jaynes - Cummings Model (JCM) is come to.

Notice that the antiresonant terms $g\,(\mathbf{b}^+\,\mathbf{J}_+ + \mathbf{b}\,\mathbf{J}_-)$ have appeared in similar hamiltonian for electron - phonon interaction in solids. The Jaynes - Cummings Model without RWA is called in the literature [27] the "dressed" JCM.

As an evident generalization of "dressed" JCM, we may consider first the hamiltonian of a $(2\,j+1)-$ level atom (spin $j = 1/2, 1, \ldots$) in the form of Eq.(2.1) with the $(2\,j+1)-$ dimensional generators of $SU(2)$ group. Secondly, in many cases $m-$ quantum transitions between atomic energy levels are of present interest. It leads to nonlinear in bosonic operators generalizations of interaction hamiltonian

$$\mathbf{H}^{(1)}_{int} = g\,(\mathbf{b})^m\mathbf{J}_+ + g\,(\mathbf{b}^+)^m\mathbf{J}_+ + h.c.. \tag{2.6}$$

It is important to notice for the future considerations that all these hamiltonians can be expressed as an operator valued functions of the direct product of Lie groups $W(1) \otimes SU(2)$ generators (the first group corresponds to field (photon) subsystem and the second - to atomic subsystem.)

The atomic subsystem's dynamical symmetry group $SU(2)$ has the non-compact group $SU(1,1)$ as an analitic continuation [26]. Unlike compact $SU(2)$ with unit sphere as quotient space, quotient space of $SU(1,1)$ is a hyperboloid. The unit circle's interior is suitable substitution for this space at complex plane, its boundary meets Lobachevsky plane infinity. Hamiltonian with generators $SU(1,1)$ instead of generators $SU(2)$ is appearing to describe three-wave parametric interaction in nonlinear media. The main difference between $SU(2)$ and $SU(1,1)$ ("hyperbolic" JC) models consist here in the role of nonresonant terms. Their presence in hamiltonian actuates a parametric build-up of oscillations and leads to energy and number of quanta increase in all subsystems. Therefore, the process requires external time dependent sources.

It is cause to write down Hamiltonian of the model as the following

$$\mathbf{H} = \nu(\mathbf{b^+b} + \frac{1}{2}) + 2\omega \mathbf{K_0} + g\,\mathbf{K_+}\mathbf{b} + \overline{g}\,\mathbf{K_-}\mathbf{b^+} + \lambda \mathbf{K_+}\mathbf{b} + \overline{\lambda}\mathbf{K_-}\mathbf{b^+}, \quad (2.7)$$

$\mathbf{K_0}$ and $\mathbf{K_\pm}$ are generators of $SU(1,1)-$ group, ( $[\mathbf{K_0},\mathbf{K_\pm}] = \pm\,\mathbf{K_\pm}$, $[\mathbf{K_+},\mathbf{K_-}] = -2\,\mathbf{K_0}$ ); $\lambda$ is (the strength of the external field - system interaction) determined by properties of external classical electromagnetic field and $\nu \approx 2\,\omega$ is treated as a condition of resonance. (Generally, $2\,\omega = \omega_1 + \omega_2$, where $\omega_{1,2}$ are the frequencies of electromagnetic waves parametrically coupled with wave of the frequency $\nu$, and constant $g$ is proportional to some component of tensor of the third order nonlinear susceptibility.)

Notice, here it is not necessary for $\lambda$ be equal of $g$, and in general, $\lambda$ is not a constant.

Considering that all introduced model hamiltonians possess dynamical symmetry group described above, coherent state technique would be applied. The technique for linear of dynamical group generators hamiltonian permits us to get explicit equations for coherent state parameters' motion, unlike in our case approximate equations are expected only. In our approach coherent state means the direct product of group - theoretical coherent states of $W(1)$, and "atomic" dynamical group ($SU(2)$ or $SU(1,1)$):

$$|CS> = |\alpha> \otimes |\zeta>, \tag{2.8}$$

here $|\alpha>$ is usual Glauber CS and $|\zeta>$ is "atomic" CS [26].

### III. COHERENT STATES, PATH INTEGRALS AND EQUATIONS OF MOTION

The purpose of this section is to get a semiclassical equations of motion in coherent state representation for JCM and its generalizations. Most natural way to do that consists in using a path integral formulation of the problem [25,28].

Consider a quantum system with the hamiltonian being an operator - valued function of the generators of the dynamical group $W(1) \otimes SU(2)$ ($SU(1,1)$) and use Schweber's Method. Here it is described briefly. Introduce a matrix element of the evolution operator of a system $U(t, t_0) = \exp(-i(t-t_0)\mathbf{H})$, between group - theoretical coherent states

$$<CS|exp(-i(t-t_0)\mathbf{H}|CS'> =$$

$$= \underbrace{\int \cdots \int}_{(N-1)} <CS_N|\exp(-i(t-t_0)\mathbf{H}/N)|CS_{N-1}> d\mu(CS_{N-1}) \times$$

$$<CS_{N-1}|\exp(-i(t-t_0)\mathbf{H}/N)|CS_{N-2}> d\mu(CS_{N-2}) \times \ldots$$

$$\ldots \times d\mu(CS_1) <CS_1|\exp(-i(t-t_0)\mathbf{H}/N)|CS_0>, \tag{3.1}$$

$$\Delta t = \frac{t-t_0}{N}, \quad N \gg 1,$$

$d\mu(CS_i)$ is the invariant measure in the resolution of unity. Here group property of evolution operator and completeness condition are involved. Each factor in Eq.( 3.1) is transforming:

$$<CS_{j+1}|e^{-i(t-t_0)\mathbf{H}/N}|CS_j> \simeq <CS_{j+1}|1 - i(t-t_0)\mathbf{H}/N|CS_j> \simeq$$

$$\simeq <CS_{j+1}|CSj> \exp(-i(t-t_0)\mathrm{h}\,(CS_{j+1}|CS_j)),$$

$$\mathrm{h}\,(KC_{j+1}|KC_j) \equiv \frac{<CS_{j+1}|\mathbf{H}|CS_j>}{<CS_{j+1}|CS_j>} \equiv \mathrm{h}(\alpha,\zeta|\alpha,\zeta)$$

is a covariant symbol of Hamiltonian [29].

Then we go over to limit by $N \to \infty$ and have as result:

$$<\alpha,\zeta|\exp(-i(t-t_0)\mathbf{H})|\alpha\prime,\zeta\prime> = \int DM_\infty\, e^{i\mathcal{S}}\quad,$$

$\mathcal{S}$ - is treated as action function, and Euler - Lagrange equations take the following form after some algebra for system with symmetry group $W(1) \otimes SU(2)$ :

$$\begin{cases} \dot{\alpha} + i\nu\alpha + i\,2jgm(\alpha^*)^{m-1}\frac{\zeta}{1+\zeta\zeta^*} + i\,2j\lambda m(\alpha^*)^{m-1}\frac{\zeta^*}{1+\zeta\zeta^*} = 0 \\ \\ \dot{\alpha}^* - i\nu\alpha^* - i\,2j\bar{g}m(\alpha)^{m-1}\frac{\zeta^*}{1+\zeta\zeta^*} - i\,2j\bar{\lambda}m(\alpha)^{m-1}\frac{\zeta}{1+\zeta\zeta^*} = 0 \\ \\ \dot{\zeta} + i\omega\zeta + i\,\bar{g}(\alpha)^m + i\,\lambda(\alpha^*)^m - i\,g(\alpha^*)^m\zeta^2 - i\,\bar{\lambda}(\alpha)^m\zeta^2 = 0 \\ \\ \dot{\zeta}^* - i\omega\zeta^* - i\,g(\alpha^*)^m - i\,\bar{\lambda}(\alpha)^m + i\,\bar{g}(\alpha)^m\zeta^{*\,2} + i\,\lambda(\alpha^*)^m\zeta^{*\,2} = 0, \end{cases} \quad (3.2)$$

here $(2j+1)$ is the number of atomic levels.

Note, so symbols $^*$ as $^-$ designate complex conjugation, the first for coherent-state parametres and the second for coupling costants. In the following calculations we are considering two particular cases: $g \neq 0, \lambda = 0$ - (JCM), and $g = \lambda$ - ("dressed" JCM).

Similar equations are led to system with dynamical group $W(1) \otimes SU(1,1)$ :

$$\begin{cases} \dot{\alpha} + i\nu\alpha + i\,2kgm(\alpha^*)^{m-1}\frac{\zeta}{1-\zeta\zeta^*} + i\,2k\lambda m(\alpha^*)^{m-1}\frac{\zeta^*}{1-\zeta\zeta^*} = 0 \\ \\ \dot{\alpha}^* - i\nu\alpha - i\,2k\bar{g}m(\alpha)^{m-1}\frac{\zeta^*}{1-\zeta\zeta^*} - i\,2k\bar{\lambda}m(\alpha)^{m-1}\frac{\zeta}{1-\zeta\zeta^*} = 0 \\ \\ \dot{\zeta} + i\omega\zeta + i\,\bar{g}(\alpha)^m + i\,\lambda(\alpha^*)^m + i\,g(\alpha^*)^m\zeta^2 + i\,\bar{\lambda}(\alpha)^m\zeta^2 = 0 \\ \\ \dot{\zeta}^* - i\omega\zeta^* - i\,g(\alpha^*)^m - i\,\bar{\lambda}(\alpha)^m - i\,\bar{g}(\alpha)^m\zeta^{*\,2} - i\,\lambda(\alpha^*)^m\zeta^{*\,2} = 0 \end{cases} \quad (3.3)$$

Instead of $j$ here arises the parameter $k$ related with eigenvalue of the $SU(1,1)-$ group Casimir operator $\mathbf{K}^2 = \mathbf{K}_0^2 - 1/2\,(\,\mathbf{K}_+\mathbf{K}_- + \mathbf{K}_-\mathbf{K}_+\,)$. For two-mode oscillator $k = \frac{1+|\Delta n|}{2} = 1/2, 1, 3/2, \ldots$ ($\Delta n$ is the difference between the number of quanta in modes); in one-mode oscillator with two-quantum transitions two possible values exist: $k = 1/4$ for even levels, and $3/4$ for odd levels.

These equations are non-linear ones and have explicit analytical solutions in some trivial cases only. Therefore, they should be solved by computer calculations.

Using Dormand - Prince Method of fifth order computer program permits us to get graphics of real and imaginary components of coherent states parametres' time evolution, complex - plane evolution of CS - parametres both for field (in $\alpha$ plane and for "atom" (in the complex plane realizations of quotient spases of dynamical groups $SU(2)$ and $SU(1,1)$. Evolution of atomic ($SU(2)$) coherent state on the Bloch-sphere is also realized when the complex parameter $\zeta$ is replaced by two angle parameters with the substitution: $\zeta = e^{i\phi}\cot(\frac{\theta}{2})$ that gives a stereographic map of a complex plane to unit sphere [26] with the South Pole corresponding to the origin of the plane. Moreover, phase portraits of real and imaginary coherent state parameter components and the transition probability time-dependence are received. For definiteness we restrict in the paper the consideration of the case to find atom in "down" state:

$$P(t) = |<down|\zeta(t)>|^2. \qquad (3.4)$$

## IV. RESULTS OF COMPUTER SIMULATIONS

In present section the results of computer calculations are described.

It is well known, Jaynes - Cummings Model has regular dynamics, for example, up-down transition probability's behaviour obeys formula [10]

$$p_{+-} = \frac{q^2(n+1)}{(\nu-\omega)^2 + (n+1)q^2}\sin^2\left[\sqrt{(\nu-\omega)^2 + q^2(n+1)}\frac{(t-t_0)}{2}\right], \qquad (4.1)$$

if number of photons in the initial state is fixed. Analogous but more unwieldy formulae exist for some other type initial states, including coherent state.

However, calculation with non-resonance terms leads to considerable complication of time-dependence behaviour so transitions probability as coherent-state parameters. This complication is most conspicuous in case of a small average number of photons ($<n>\sim 1$) in the initial state of field mode and large coupling constants. The following regularity might be noted: while $g \leq 0.05 \div 0.1$ (magnitude of coupling constant depends on initial values of atomic and field coherent - state parameters) transitions probability time behaviour remains "sine-similar" curve, then with increasing $g$ "sine-similar" envelope of curve exists only and multifrequency regime is observed. Further, if coupling constant is continuing its growth, there is no harmonical envelope of the curve, but such value of $g$ exists when number of probability oscillations frequency cuts down (to two frequencies, in our case) (as Fig.1, which has been calculated with $g = \lambda = 0.5$, $\nu = \omega = 1$,, and system has been prepared in the initial state: $|\alpha(0)> \otimes |\zeta(0)>$ , $\zeta(0) = 0.3$, $\alpha(0) = 1$.). After this, system (in sense of probability behaviour) returns to multifrequency regime (Figs.2, 3) and draws near the chaotic regime (Fig.4). The effect takes place with $<n>$ essentially more then 1 (about 25, in case is considered), what have been described. The effect is observed on lesser value of the coeffitient $g$ then it is necessary for small average number of photons in the initial field mode. It is important then $<n> \gg 1$ corresponds to classical field and the problem reduces itself to problem of "spin dynamics in the (magnetical) field", with no chaotic solutions. The field coherent state point is interested to move over constant radius circle with very small distorsions. Also, case of two - quantum ($2\nu = \omega = 1$) transitions "dressed" JCM have been investigated. Some typical results are showed in Figs.5 - 9 (here $g = \lambda = 0.4$, $\zeta(0) = 0.3$). The first picture ($\alpha(0) = 1$) of them presents chaotic behaviour of system unlike second one with larger initial field coherent state value ($\alpha(0) = 5$) is multifrequence regime. These conclusions are founded by observing of calculating phase portraits (see Figs.7, 8). The last picture (Fig.9) in series demonstrates the beginning evolution coherent state point over Bloch - sphere.

In conclusion, systems behaviour in both described cases are similar, but in two-quantum "dressed" JCM a chaotic regime arises with in order smaller of coupling constant magnitude. Although, we guess that needed value achievement in real experiment is very difficult problem.

The possibility of "dressed" JCM chaotic behaviour is provided by antiresonant terms, which might describe as single virtual two-photon process. Similar terms may be included in hamiltonian of three-wave parametric interaction, but their source must be found out of system, because energy conservation law (for closed system) is violated by the terms. The medium with non-linearity of forth order is occurred the cause of parametric build-up process, for example. The short-time exposure is simulated by step-function $\lambda(t)$ : this function identically equals zero everywhere except short (as compared with evolution time) interval. It is occured, this way does not lead to chaos. The changes in phase portraits are observed, and there is a marked difference between some of portrait, but all of them stay as regular structures (Figs. 10 - 13). The atomic coherent state point approaches to bound of unit disc for the finite time, and the time decreases with increasing coupling constant or interaction time interval.

The same effects are observed for time - dependence of coherent state variables. For example, the Figs.14, 15 show dependence $Re\,\zeta$, when activating the stepwise parametric interaction with following parameters of system: $\nu = 2\,\omega = 1$, $g = 0.2$, for initial state determined by $\zeta(0) = 0.5$, $\alpha(0) = 5$; the interaction was switched after $\omega\,t_0 = 10$ free evolution of system for interval $\delta(\omega t_0) = 0.2$, $\lambda = 0.2$.

The destruction of system is a possible result of exposure parameters increasing. Physical treatment of this is exponential growth of quanta average number in so field as "atomic" modes. As result, hyperbolic JCM is simpler in "chaotic sense". Perhaps, the behaviour of system nearly parametric generation domain requires detailed studying by more accurated computation procedure unlike used.

## V. CONCLUSIONS

In the paper we restrict our discussion only particular cases of the models (two - level atom ($j = 1/2$) or degenerate "atomic" system with dynamical symmetry group $SU(1,1)$ ($k = 1/4$ or $k = 3/4$)). The computer program allows to do similar calculations for wide class of possible evident model enlargement.

Besides, the developed approach gives a natural way to its generalization for chaotic quantum corrections' calculations, and presented results might be consider as the first step of our investigation of chaos in quantum optical models.

Method of Heisenberg operator averages, based on similar ideas (but non technique), gives the opportunity to involve wider class of problems because it don't needs of assumption about conservation subsystems' states as coherent during the evolution. Nevertheless, account of quantum corrections is believed to be more complicated technically in this approach.

We compared the results in both approaches for "dressed" Jaynes - Cummings Model by calculating inverted population difference. All of them are in good agreement. Comparison of these methods and computer simulations for some extension of quantum optical model hamiltonians will be published in the next paper.

## ACKNOWLEDGMENTS

We are grateful to Dr. E.K.Bashkirov, Dr. A.A.Biryukov, and Dr.V.L.Derbov for helpful and stimulating discussions.

This work was partially supported by the International Scientific Foundation, grant No. J 64100.

# Figure Captions

**Fig.1.** Time dependence of the down - state population in two - level atom with $g = \lambda = 0.5$, $\nu = \omega = 1$, $\zeta(0) = 0.3$, $\alpha(0) = 1$.

**Fig.2.** Time dependence of the down - state population in two - level atom with $g = \lambda = 0.6$, $\nu = \omega = 1$, $\zeta(0) = 0.3$, $\alpha(0) = 1$.

**Fig.3.** Time dependence of the down - state population in two - level atom with $g = \lambda = 0.7$, $\nu = \omega = 1$, $\zeta(0) = 0.3$, $\alpha(0) = 1$.

**Fig.4.** Time dependence of the down - state population in two - level atom with $g = \lambda = 0.8$, $\nu = \omega = 1$, $\zeta(0) = 0.3$, $\alpha(0) = 1$.

**Fig.5.** Time dependence of the down - state population with the two-photon transition in two - level atom and $g = \lambda = 0.4$, $2\nu = \omega = 1$, $\zeta(0) = 0.3$, $\alpha(0) = 1$.

**Fig.6.** Time dependence of the down - state population with the two-photon transition in two - level atom and $g = \lambda = 0.4$, $2\nu = \omega = 1$, $\zeta(0) = 0.3$, $\alpha(0) = 5$.

**Fig.7.** Phase portrait in the plane $(Re\,\zeta; Re\dot{\zeta})$ ($g = \lambda = 0.4, 2\nu = \omega = 1, \alpha(0) = 1, \zeta(0) = 0.3$.)

**Fig.8.** Phase portrait in plane $(Re\,\zeta; Re\dot{\zeta})$ ($g = \lambda = 0.4, 2\nu = \omega = 1, \alpha(0) = 5, \zeta(0) = 0.3$).

**Fig.9.** Bloch - sphere atomic coherent state evolution ($g = \lambda = 0.4, 2\nu = \omega = 1, \alpha(0) = 1, \zeta(0) = 0.3$).

**Fig.10.** $SU(1,1)-$ coherent state dynamics on unit disc $|\zeta| < 1$, when activating the stepwise parametric interaction ($k = 1/4, \nu = 2\omega = 1, g = 0.2, \alpha(0) = 1, \zeta(0) = 0.5; \delta(\omega t_0) = 2, \omega\,t_0 = 10, \lambda = 0.5$).

**Fig.11.** Phase portrait in plane $(Re\,\zeta; Re\dot{\zeta})$ when activating the stepwise parametric interaction ($k = 1/4, \nu = 2\omega = 1, g = 0.2, \alpha(0) = 1, \zeta(0) = 0.5; \delta(\omega t_0) = 2, \omega\,t_0 = 10, \lambda = 0.5$).

**Fig.12.** $SU(1,1)-$ coherent state dynamics on unit disc $|\zeta| < 1$, when activating the stepwise parametric interaction ($k = 1/4, \nu = 2\omega = 1, g = 0.2, \alpha(0) = 1, \zeta(0) = 0.5; \delta(\omega t_0) = 3, \omega\,t_0 = 10, \lambda = 0.5$).

**Fig.13.** Phase portrait in plane $(Re\,\zeta; Re\,\dot\zeta)$ when activating the stepwise parametric interaction ($k = 1/4, \nu = 2\,\omega = 1, g = 0.2, \alpha(0) = 1, \zeta(0) = 0.5; \delta(\omega t_0) = 3, \omega\,t_0 = 10, \lambda = 0.5$).

**Fig.14.** Time dependence $Re\,\zeta$, when activating the stepwise parametric interaction ($k = 1/4, \nu = 2\,\omega = 1, g = 0.2, \alpha(0) = 5, \zeta(0) = 0.5; \delta(\omega t_0) = 0.2, \omega\,t_0 = 10, \lambda = 0.2$).

**Fig.15.** Time dependence $Re\,\zeta$, when activating the stepwise parametric interaction ($k = 1/4, \nu = 2\,\omega = 1, g = 0.2, \alpha(0) = 5, \zeta(0) = 0.5; \delta(\omega t_0) = 0.2, \omega\,t_0 = 10, \lambda = 0.2$.) (The initial part of **Fig.14.**, $\omega\,t \leq 40.0$).